\documentclass{article}
\usepackage{ijcai22}

\usepackage{times}
\usepackage{soul}
\usepackage{url}
\usepackage[dvipsnames]{xcolor}
\usepackage[colorlinks=true,linkcolor=Blue, citecolor=Blue, filecolor=magenta, urlcolor=violet]{hyperref}

\usepackage[utf8]{inputenc}
\usepackage[small]{caption}
\usepackage{graphicx}
\usepackage{subcaption}
\usepackage{booktabs}
\usepackage{multirow}

\usepackage{array}
\newcolumntype{H}{>{\setbox0=\hbox\bgroup}c<{\egroup}@{}}

\newcommand{\proj}{\text{Proj}}
\newcommand{\vdelta}{\bm \delta}

\newcommand{\tit}[1]{\textit{#1}}
\newcommand{\tbf}[1]{\textbf{#1}}

\newcommand{\thl}[1]{\tbf{#1}} % for highlight
\newcommand{\tHL}[1]{\underline{#1}} % for highlight

\newcommand{\etal}{\emph{et al. }}

\usepackage{amsmath,amsfonts, amssymb, mathtools, bm, bbm, nicefrac}
\newcommand{\mcal}[1]{\mathcal{#1}}

\usepackage{algorithm}
\usepackage[noend]{algpseudocode}

\newcommand{\E}{\mathbb{E}}

\DeclareMathOperator*{\argmax}{arg\,max}

\def\sR{{\mathbb{R}}}

\def\rz{{\textnormal{z}}}

\def\rvv{{\mathbf{V}}}

\def\rvx{{\mathbf{X}}}
\def\rvy{{\mathbf{Y}}}

\def\vn{{\mathbf{n}}}

\def\vv{{\mathbf{v}}}

\def\vx{{\mathbf{x}}}
\def\vy{{\mathbf{y}}}
\def\vz{{\mathbf{z}}}

\newtheorem{theorem}{Theorem}

\title{Towards Adversarially Robust Deep Image Denoising}
 \iftrue
\author{
	Hanshu Yan$^1$\and
	Jingfeng Zhang$^2$\and
	Jiashi Feng$^3$\and
	Masashi Sugiyama$^{2,4}$\and
	Vincent Y. F. Tan$^{1,5}$
	\affiliations
	$^1$ECE, NUS;
	$^2$RIKEN-AIP;
	$^3$ByteDance Inc;
	$^4$GSFS, UTokyo;
	$^5$Math, NUS
	\emails
	hanshu.yan@u.nus.edu
}
\fi

\begin{document}
\maketitle
\begin{abstract}
This work systematically investigates the adversarial robustness of deep image denoisers (DIDs), i.e, how well DIDs can recover the ground truth from noisy observations degraded by adversarial perturbations. Firstly, to evaluate DIDs' robustness,  we propose a novel adversarial attack, namely Observation-based Zero-mean Attack ({\sc ObsAtk}), to craft adversarial zero-mean perturbations on given noisy images. We find that existing DIDs are vulnerable to the adversarial noise generated by {\sc ObsAtk}. Secondly, to robustify DIDs, we propose an adversarial training strategy, hybrid adversarial training ({\sc HAT}), that jointly trains DIDs with adversarial and non-adversarial noisy data to ensure that the reconstruction quality is high and the denoisers around non-adversarial data are locally smooth. The resultant DIDs can effectively remove various types of synthetic and adversarial noise. We also uncover that the robustness of DIDs benefits their generalization capability on unseen real-world noise. Indeed, {\sc HAT}-trained DIDs can recover high-quality clean images from real-world noise even without training on real noisy data. Extensive experiments on benchmark datasets, including Set68, PolyU, and SIDD, corroborate the effectiveness of {\sc ObsAtk} and {\sc HAT}.
\end{abstract}

\section{Introduction}
Image denoising, which aims to reconstruct clean images from their noisy observations, is a vital part of the image processing systems. The noisy observations are usually modeled as the addition between ground-truth images and zero-mean noise maps \cite{dabov_image_2007,zhang_beyond_2017}. Recently, deep learning-based methods have made significant advancements in denoising tasks \cite{zhang_beyond_2017,anwar_real_2019} and have been applied in many areas including medical imaging \cite{gondara_medical_2016} and photography \cite{abdelhamed_high-quality_2018}. Despite the success of deep denoisers in recovering high-quality images from a certain type of noisy images, we still lack knowledge about their robustness against adversarial perturbations, which may cause severe safety hazards in high-stake applications like medical diagnosis. To address this problem, the first step should be developing attack methods dedicated for denoising to evaluate the robustness of denoisers. In contrast to the attacks for classification \cite{goodfellow_explaining_2015,madry_towards_2018}, attacks for denoising should consider not only the adversarial budget but also some assumptions of natural noise, such as zero-mean, because certain perturbations, such as adding a constant value, do not necessarily result in visual artifacts. Although Choi \etal \shortcite{choi_deep_2021,choi_evaluating_2019} studied the vulnerability for various deep image processing models, they directly applied the attack from classification. To the best of our knowledge, no attacks are truly dedicated for the denoising task till now.

To this end, we propose the observation-based zero-mean attack ({\sc ObsAtk}), which crafts a worst-case zero-mean perturbation for a noisy observation by maximizing the distance between the output and the ground-truth. To ensure that the perturbation satisfies the adversarial budget and the zero-mean constraints, we utilize the classical projected-gradient-descent (PGD) \cite{madry_towards_2018} method for optimization, and develop a two-step operation to project the perturbation back into the feasible region. Specifically, in each iteration, we first project the perturbation onto the zero-mean hyperplane. Then, we linearly rescale the perturbation to adjust its norm to be less or equal to the adversarial budget.  We examine the effectiveness of {\sc ObsAtk} on several benchmark datasets and find that deep image denoisers are indeed susceptible to {\sc ObsAtk}: the denoisers cannot remove adversarial noise completely and even yield atypical artifacts, as shown in Figure \ref{fig:adv-noise-advout}.

To robustify deep denoisers against adversarial perturbations, we propose an effective adversarial training strategy, namely hybrid adversarial training ({\sc HAT}), to train denoisers by using adversarially noisy images and non-adversarial noisy images together. The loss function of {\sc HAT} consists of two terms. The first term ensures the reconstruction performance from common non-adversarial noisy images, and the second term ensures the reconstructions between non-adversarial and adversarial images to be close to each other. Thus, we can obtain denoisers that perform well on both non-adversarial noisy images and their adversarial perturbed versions. Extensive experiments on benchmark datasets verify the effectiveness of {\sc HAT}. 

Moreover, we reveal that adversarial robustness benefits the generalization capability to unseen types of noise, i.e., {\sc HAT} can train denoisers for real-world noise removal only with synthetic noise sampled from common distributions like Gaussians. That is because {\sc ObsAtk} searches for the worst-case perturbations around different levels of noisy images, and training with adversarial data ensures the denoising performance on various types of noise. In contrast, other reasonable methods for real-world denoising \cite{guo_toward_2019,lehtinen_noise2noise_2018} mostly require a large number of real-world noisy data for the training, which are unfortunately not available in some applications like medical radiology. We conduct experiments on several real-world datasets. Numerical and visual results demonstrate the effectiveness of {\sc HAT} for real-world noise removal.

In summary, there are three main contributions in this work: 1) We propose a novel attack, {\sc ObsAtk}, to generate adversarial examples for noisy observations, which facilitates the evaluation of the robustness of deep image denoisers. 2) We propose an effective adversarial training strategy, {\sc HAT}, for robustifying deep image denoisers. 3) We build a connection between adversarial robustness and the generalization to unseen noise, and show that {\sc HAT}   serves as a promising framework for training generalizable deep image denoisers.

\section{Notation and Background}
\paragraph{Adversarial robustness and adversarial training}
Consider a deep neural network (DNN) $\{f_{\theta}:\theta\in \Theta\}$ mapping an input $\vy$ to a target $\vx$, the model is trained to minimize a certain loss function that is measured by particular distance $d(\cdot,\cdot)$ between output $f_{\theta}(\vy)$ and the target $\vx$. In high stake applications, the DNN should resist small perturbations on the input data and map the perturbed input to a result close to the target. The notion of \tit{robustness} has been proposed to measure the resistance of DNNs against the slight changes of the input \cite{szegedy_intriguing_2014,goodfellow_explaining_2015}. The robustness is characterized by the distance $d(f_{\theta}(\vy'),\vx)$ between $f_{\theta}(\vy')$ and target $\vx$, where the worst-case perturbed input $\vy'$ is located within a small neighborhood of the original input $\vy$ and maximizes the distance between its output and target $\vx$.
\begin{equation}
	\vy' = \argmax_{\vy':\|\vy'-\vy\|\leq \rho } d(f_{\theta}(\vy'),\vx).
	\label{eq:adv_attacks}
\end{equation}
The worst-case perturbation $\vy'$ can be approximated via many adversarial attack methods, such as FGSM \cite{goodfellow_explaining_2015}, I-FGSM \cite{kurakin_adversarial_2017}, and PGD \cite{madry_towards_2018}, which solve \eqref{eq:adv_attacks} via gradient descent methods.
The distance $d(f_{\theta}(\vy'),\vx)$ is an indication of the robustness of $f_{\theta}$ around $\vy$: a small distance implies strong robustness and vice versa.
In terms of image classification, the $\rho$-neighborhood is usually defined by the $\ell_{\infty}$-norm and the distance $d(\cdot, \cdot)$ is measured by the cross-entropy loss \cite{madry_towards_2018} or a margin loss \cite{carlini_towards_2017}. For image restoration, the distance between images is usually measured by the $\ell_2$-norm \cite{zhang_beyond_2017}. 

In most cases, deep learning models have been shown to be vulnerable against adversarial attacks under normal training (NT) \cite{tramer_adaptive_2020,yan_robustness_2019}. To robustify DNNs, Madry \etal \shortcite{madry_towards_2018} proposed the PGD adversarial training (AT) method which trains DNNs with adversarial examples of the original data. AT is formally formulated as the following min-max optimization problem,
\begin{equation}
	\min_{\theta\in \Theta} \max_{\vy':\|\vy'-\vy\|\leq \rho } d(f_{\theta}(\vy'),\vx).
	\label{eq:pgd-at}
\end{equation}
Its effectiveness has been verified by extensive empirical and theoretical results \cite{yan_cifs_2021,gao_convergence_2019}. For further improvement, many variants of PGD have been proposed in terms of its robustness enhancement \cite{zhang_theoretically_2019}, generalization to non-adversarial data \cite{zhang_attacks_2020}, and computational efficiency \cite{shafahi_adversarial_2019}.

\paragraph{Deep image denoising} During image capturing, unknown types of noise may be induced by physical sensors, data compression, and transmission. Noisy observations are usually modeled as the addition between the ground-truth images and certain zero-mean noise \cite{dabov_image_2007,zhang_poisson_gaussian_2019}, i.e., $\rvy = \rvx + \rvv$ with $\E_{Q}\big[ \sum_{i=1}^{m} \rvv_{[i]} \big] = 0$, where $\rvv_{[i]}$ is the $i^{\text{th}}$ element of $\rvv$. The random vector $\rvx \in \sR^m$ with  distribution $P$ denotes a random clean image and the noise $\rvv \in \sR^m$ with a distribution $Q$ satisfies the zero-mean constraint. 
Denoising techniques aim to recover clean images from their noisy observations \cite{zhang_beyond_2017,dabov_image_2007}. Suppose we are given a training set  $\mcal S=\{(\vy_j, \vx_j)\}^{N}_{j=1}$ of noisy and clean image pairs sampled from distributions $Q$ and $P$ respectively, we can train a DNN to effectively remove the noise induced by distribution $Q$ from the noisy observations. A series of DNNs have been developed for denoising in recent years, including DnCNN~\cite{zhang_beyond_2017}, FFDNet~\cite{zhang_ffdnet_2018}, and RIDNet~\cite{anwar_real_2019}.

In real-world applications \cite{abdelhamed_high-quality_2018,xu_multi-channel_2017}, the noise distribution $Q$ is usually unknown due to the complexity of the image capturing procedures; besides, collecting a large number of image pairs (clean/noisy or noisy/noisy) for training sometimes may be unrealistic in safety-critical domains such as medical radiology \cite{zhang_poisson_gaussian_2019}. To overcome these, researchers developed denoising techniques by approximating real noise with common distributions like Gaussian or Poisson \cite{dabov_image_2007,zhang_poisson_gaussian_2019}. 
To train denoisers that can deal with different levels of noise, where the noise level is measured by the energy-density $\|\vv\|^2_2/m$ of noise, the training set may consist of noisy images sampled from a variety of noise distributions \cite{zhang_beyond_2017}, whose expected energy-densities range from zero to certain budget $\epsilon^2$ (the expected $\ell_2$-norms range from zero to $\epsilon \sqrt{m}$).
For example, $\mcal S^{\epsilon}=\{(\vy_j, \vx_j)\}^{N}_{j=1}$  where $\vy_j=\vx_j+\vv_j$ and $\vx_j$ and $\vv_j$ are sampled from $P$ and $Q$ respectively and where $Q$ is randomly selected from a set of Gaussian distributions 
$\mcal Q^\epsilon = \{ \mcal N(\bm{0}, \sigma^2 \mathbf{I})|   \sigma\in [0, \epsilon] \}$. 
The denoiser $f_{\theta}^{\epsilon}(\cdot)$ trained with $\mcal S^{\epsilon}$ is termed as an $\epsilon$-denoiser.

\paragraph{On robustness of deep image denoisers}
In practice, data storage and transmission may induce imperceptible perturbations on the original data so that the perturbed noise may be statistically slightly different from the noise sampled from the specific original distribution.
Although an $\epsilon$-denoiser can successfully remove noise sampled from $Q \in \mcal Q^{\epsilon}$, the performance of noise removal on the perturbed data is not guaranteed. Thus, we propose a novel attack method, {\sc ObsAtk}, to assess the adversarial robustness of DIDs in Section \ref{sec:obsattack}. To robustify DIDs, we propose an adversarial training strategy, {\sc HAT}, in Section \ref{sec:hat}. {\sc HAT}-trained DIDs can effectively denoise adversarial perturbed noisy images and preserve good performance on non-adversarial data. 

Besides the adversarial robustness issue, it has been shown that $\epsilon$-denoisers trained with $\mcal S^{\epsilon}$ cannot generalize well to unseen real-world noise \cite{lehtinen_noise2noise_2018,batson_noise2self_2019}. 
Several methods have been proposed for real-world noise removal, but most of them require a large number of real noisy data for training, e.g., CBDNet (clean/noisy pairs) \cite{guo_toward_2019} and Noise2Noise (noisy pairs) \cite{lehtinen_noise2noise_2018}, which is sometimes impractical. In Section \ref{sec:hat_for_unseen}, we show that {\sc HAT}-trained DIDs can generalize well to unseen real noise without the need of utilizing real noisy images for training.

\section{{\sc ObsAtk} for Robustness Evaluation}
\label{sec:obsattack}
In this section, we propose a novel adversarial attack, Observation-based Zero-mean Attack ({\sc ObsAtk}), to evaluate the robustness of DIDs. We also conduct experiments on benchmark datasets to demonstrate that normally-trained DIDs are vulnerable to adversarial perturbations.

\subsection{Observation-based Zero-mean Attack}
\begin{figure}[t!]
	\centering
	\includegraphics[width=.8\linewidth]{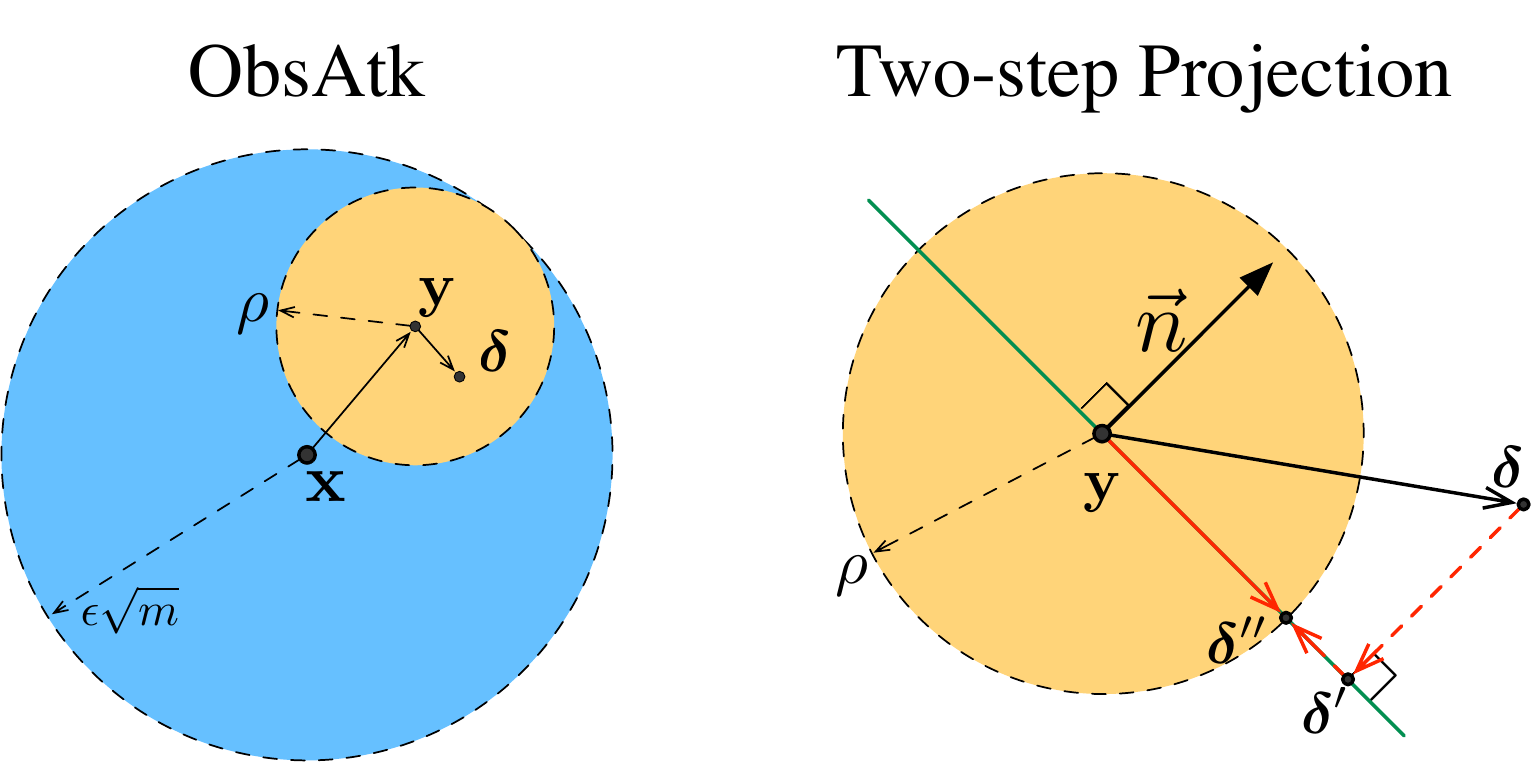}
	\caption{Illustration of {\sc ObsAtk}. Left: We perturb a noisy observation $\vy$ of the ground-truth $\vx$ with an adversarial budget $\rho$ in the $\ell_2$-norm. For an $\epsilon$-denoiser, we choose a proper value of $\rho$ to ensure the norm of the total noise is bounded by $\epsilon\sqrt{m}$, where $m$ denotes the image size. Right: The perturbation $\vdelta$ is projected via the two-step operation onto the region defined by the zero-mean and $\rho$-ball constraints.}
	\label{fig:obsattack}
\end{figure}
An $\epsilon$-denoiser $f^{\epsilon}_{\theta}(\cdot)$ can generate a high-quality reconstruction $f^{\epsilon}_{\theta}(\vy)$ close to the ground-truth $\vx$ from a noisy observation $\vy=\vx+\vv$. To evaluate the robustness of $f^{\epsilon}_{\theta}(\cdot)$ with respect to a perturbation on $\vy$, we develop an attack to search for the worst perturbation $\bm{\delta}^*$ that degrades the recovered image $f^{\epsilon}_{\theta}(\vy+\bm{\delta}^*)$ as much as possible. Formally, we need to solve the problem stated in Eq.~\eqref{eq:obsattack}. The optimization problem is subject to \tit{two} constraints: The first constraint requires the norm of $\bm{\delta}$ to be bounded by a small adversarial budget $\rho$.
The second constraint restricts the mean $M(\bm{\delta})$ of all elements in $\bm{\delta}$ to be zero.
This corresponds to the zero-mean assumption of noise in real-world applications because a small mean-shift does not necessarily result in visual noise. For example, a mean-shift in gray-scale images implies a slight change of brightness. Since the zero-mean perturbation is added to a noisy observation $\vy$, we term the proposed attack as Observation-based Zero-mean Attack ({\sc ObsAtk}).
\begin{subequations}
\begin{align}
	\bm{\delta}^* & = \argmax_{\bm{\delta} \in \sR^m} \| f_{\theta}^{\epsilon}(\vy+\bm{\delta}) - \vx\|_2^2 , \\
	\text{s.t.~~} &   \|\bm{\delta}\|_2 \leq \rho, \;\;\text{~}\;\; M(\bm{\delta})=\frac{1}{m}\sum_{i=1}^{m} \bm{\delta}_{[i]}=0.\label{eq:obsattack-constraints}
\end{align}
\label{eq:obsattack}
\end{subequations}

\begin{algorithm}[t!]
% \caption{Observation-based Zero-mean Attack $\!$ ({\sc ObsAtk})$\!$}
\caption{{\sc ObsAtk}}
\label{alg:obsattack}
\begin{algorithmic}[1]
\Require Denoiser $f_{\theta}(\cdot)$, ground-truth $\vx$, noisy observation $\vy$, adversarial budget $\rho$, \#iterations $T$, step-size $\eta$, minimum pixel value $p_{\text{min}}$, maximum pixel value $p_{\text{max}}$
\Ensure Adversarial perturbation $\bm{\delta}$
\State $\bm{\delta} \leftarrow \mathbf{0}$
	\For{$t=1$ to $T$}
	\State $\bm{\delta} \leftarrow \bm{\delta} + \eta\nabla_{\bm{\delta}}\| f_{\theta}^{\epsilon}(\vy+\bm{\delta}) - \vx\|_2^2$; 
	\State $\bm{\delta} \leftarrow \bm{\delta} - (\bm{\delta}^\top \vn /{\|{\vn}\|^2_2}) {\vn}$ where $\vn$ is in \eqref{eq:zero-mean-proj}
	\State $\bm{\delta} \leftarrow \min(\nicefrac{\rho}{\|\bm{\delta}\|_2}, 1) \bm{\delta}$;
    \EndFor
\State $\bm{\delta} \leftarrow \text{Clip}(\vy+\bm{\delta}, p_{\text{min}}, p_{\text{max}}) - \vy$
\end{algorithmic}
\end{algorithm}
\vspace{-.1in}

We solve the constrained optimization problem Eq.~\eqref{eq:obsattack} by using the classical projected-gradient-descent (PGD) method. PGD-like methods update optimization variables iteratively via gradient descent and ensure the constraints to be satisfied by projecting parameters back to the feasible region at the end of each iteration. To deal with the $\ell_2$-norm and zero-mean constraints, we develop a two-step operation in Eq.~\eqref{eq:two-step-projection}, that first projects the perturbation $\bm{\delta}$ back to the zero-mean hyperplane and then projects the result onto the $\rho$-neighborhood.
\begin{subequations}
\begin{align}
	\bm{\delta}' &= \bm{\delta} - \frac{\bm{\delta}^\top \vn}{\|{\vn}\|^2_2} {\vn},\quad\mbox{where}\quad \vn = [1,1, \ldots, 1]^\top, \label{eq:zero-mean-proj} \\
	\bm{\delta}'' &= \min\Big(\frac{\rho}{\|\bm{\delta}'\|_2}, 1\Big) \text{~} \bm{\delta}'. \label{eq:proj-norm}
\end{align}
\label{eq:two-step-projection}
\end{subequations}

In each iteration, as shown in Figure~\ref{fig:obsattack}, the first step involves  projecting the perturbation $\bm{\delta}$ onto the zero-mean hyperplane. The zero-mean hyperplane consists of all the vectors $\vz$ whose mean of all elements equals zero, i.e., $\vn^{\top} \vz=0$, where $\vn$ is the length-$d$ all ones vector. Thus, $\vn$ is a normal of the zero-mean plane. We can project any vector onto the zero-mean plane via \eqref{eq:zero-mean-proj}. The vector $\bm{\delta}$ is first projected along the direction of $\vn$, then its projection $\bm{\delta}'$ onto the zero-mean plane equals itself minus its projection onto $\vn$.
The second step involves further projecting $\bm{\delta}'$ back to the $\rho$-ball via linear scaling. If  $\bm{\delta}'$ is already within the $\rho$-ball, we keep $\bm{\delta}'$ unchanged. Otherwise, the final projection $\bm{\delta}''$ is obtained by scaling $\bm{\delta}'$ with a factor ${\rho}/{\|\bm{\delta}'\|_2}$. For any two sets $A$ and $B$, although the projection onto $A\cap B$ is, in general, not equal to the result obtained by first projecting onto $A$, then onto $B$, surprisingly, the following holds for the two sets  in \eqref{eq:obsattack-constraints}.

\begin{theorem}[Informal]
 Given any vector $\bm{\delta} \in \sR^m$, the projection of $\bm{\delta} $ via the two-step operation in \eqref{eq:two-step-projection} satisfies the two constraints in \eqref{eq:obsattack-constraints}, and the two-step projection is equivalent to the exact projection onto the set defined by \eqref{eq:obsattack-constraints}.
	\label{thm:two-step-proj}
\end{theorem}
The formal statement and the proof of Theorem \ref{thm:two-step-proj} are provided in Appendix \ref{sec:thm1-proof}. The complete procedure of {\sc ObsAtk} is summarized in Algorithm \ref{alg:obsattack}.

\def \SubFigWidth {0.13} % define a variable
\def \SubImgWidth {.95}
\begin{figure*}[t!]
    \centering
    \begin{subfigure}{\SubFigWidth\linewidth}
        \centering
        \includegraphics[width=\SubImgWidth \linewidth]{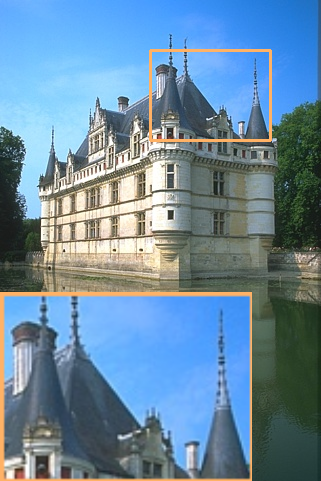}
        \caption{\scriptsize $\vx$}
    \end{subfigure}
    \begin{subfigure}{\SubFigWidth\linewidth}
        \centering
        \includegraphics[width=\SubImgWidth \linewidth]{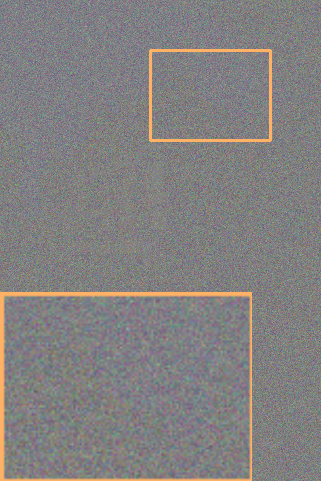}
        \caption{\scriptsize $\vv$}
    \end{subfigure}
    \begin{subfigure}{\SubFigWidth\linewidth}
        \centering
        \includegraphics[width=\SubImgWidth \linewidth]{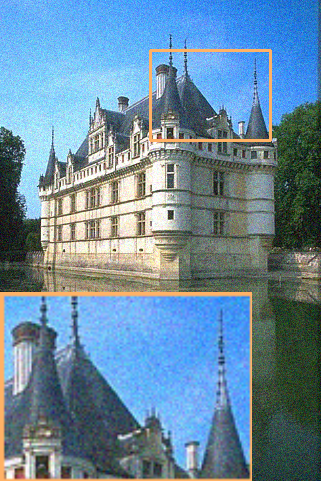}
        \caption{\scriptsize $\vy$}
    \end{subfigure}
    \begin{subfigure}{\SubFigWidth\linewidth}
        \centering
        \includegraphics[width=\SubImgWidth \linewidth]{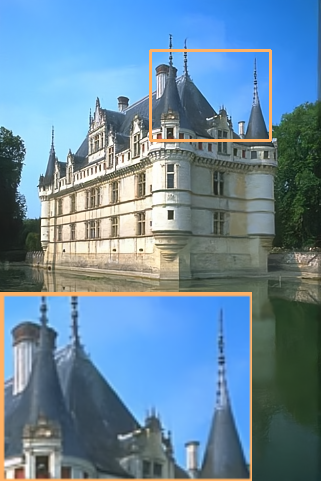}
        \caption{\scriptsize $f^{\epsilon}_{\theta}(\vy)$}
    \end{subfigure}
    \begin{subfigure}{\SubFigWidth\linewidth}
        \centering
        \includegraphics[width=\SubImgWidth \linewidth]{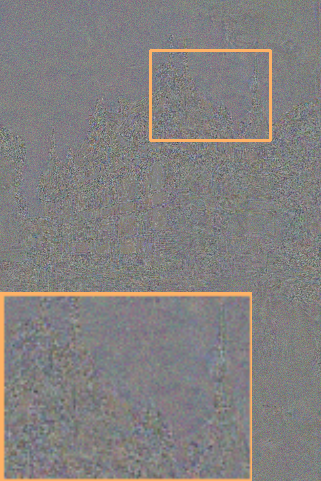}
        \caption{\scriptsize $\vv+\bm{\delta}$}
    \end{subfigure}
    \begin{subfigure}{\SubFigWidth\linewidth}
        \centering
        \includegraphics[width=\SubImgWidth \linewidth]{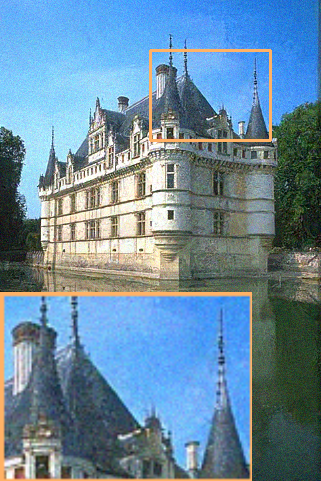}
        \caption{\scriptsize $\vy+\bm{\delta}$}
    \end{subfigure}
    \begin{subfigure}{\SubFigWidth\linewidth}
        \centering
        \includegraphics[width=\SubImgWidth \linewidth]{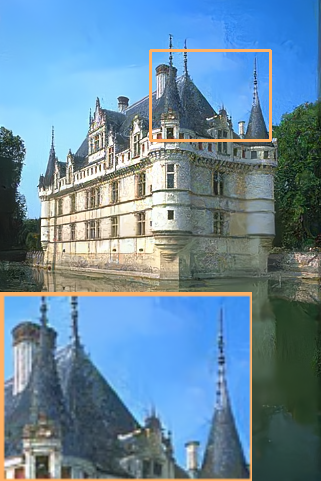}
        \caption{\scriptsize $f^{\epsilon}_{\theta}(\vy+\bm{\delta})$}
        \label{fig:adv-noise-advout}
    \end{subfigure}
	\caption{
		Given a normally-trained denoiser $f^{\epsilon}_{\theta}(\cdot)$, from left to right are the ground-truth image $\vx$, Gaussian noise $\vv$, the Gaussian noisy image $\vy=\vx+\vv$, the reconstruction  $f^{\epsilon}_{\theta}(\vy)$ from $\vy$, adversarial noise $\vv+\bm{\delta}$, the adversarially noisy image $\vy+\bm{\delta}$, and the reconstruction $f^{\epsilon}_{\theta}(\vy+\bm{\delta})$ from $\vy+\bm{\delta}$. Comparing (a), (d) and (g), we observe that $f^{\epsilon}_{\theta}(\cdot)$ can effectively remove Gaussian noise but its performance is degraded when dealing with the adversarial noise (noise remains on the roof and strange contours appear in the sky).
		} 
	\label{fig:adv-noise}
\end{figure*}

\subsection{Robustness Evaluation via {\sc ObsAtk}}

We use {\sc ObsAtk} to evaluate the adversarial robustness of $\epsilon$-denoisers on several gray-scale and RGB benchmark datasets, including Set12, Set68, BSD68, and Kodak24. For gray-scale image denoising, we use Train400 to train a DnCNN-B \cite{zhang_beyond_2017} model, which consists of 20 convolutional layers. We follow the training setting in Zhang \etal \shortcite{zhang_beyond_2017} and randomly crop $128\times 3000$ patches in size of $50\times50$. Noisy and clean image pairs are constructed by injecting different levels of white Gaussian noise into clean patches. The noise levels $\sigma$ are uniformly randomly selected from $[0, \epsilon]$ with $\epsilon=\nicefrac{25}{255}$. For RGB image denoising, we use BSD432 (BSD500 excluding images in BSD68) to train a DnCNN-C model with the same number of layers as DnCNN-B and but set the input and output channels to be three. Other settings follow those of the training of DnCNN-B. 
% Besides Gaussian denoising and DnCNN-like models, we also experiment with other DNN architectures and other types of noise. We present the results of Gaussian denoising in this section, and others are provided in Appendix \toref.

We evaluate the denoising capability of the $\epsilon$-denoiser on Gaussian noisy images and their adversarially perturbed versions. The image quality of reconstruction is measured via the peak-signal-noise ratio (PSNR) metric. A large PSNR between reconstruction and ground-truth implies a good performance of denoising. We denote the energy-density of the noise in test images as $\hat \epsilon^2$ 
and consider three levels of noise, i.e., $\hat \epsilon=\nicefrac{25}{255}$, $\nicefrac{15}{255}$, and $\nicefrac{10}{255}$.
For Gaussian noise removal, we add white Gaussian noise with $\sigma=\hat \epsilon$ to clean images. 
For Uniform noise removal, we generate noise from $\mcal U(-\sqrt{3}\hat \epsilon,\sqrt{3}\hat \epsilon)$.
For denoising adversarial noisy images, the norm budgets of adversarial perturbation are set to be $\rho=\nicefrac{5}{255}\cdot \sqrt{m}$ and $\nicefrac{7}{255}\cdot\sqrt{m}$ respectively, where $m$ equals the size of test images.
We perturb noisy observations whose noise are generated from 
$\mcal N(0,\hat \epsilon-\nicefrac{\rho}{\sqrt{m}})$, 
so that the $\ell_2$-norms of total noise in adversarial images are still bounded by $\hat \epsilon \cdot \sqrt{m}$ and the energy-density thus are bounded by $\hat \epsilon^2$ . We use Atk-$\nicefrac{\rho}{\sqrt{m}}$ to denote the adversarially perturbed noisy images in the size of $m$ with adversarial budget $\rho$. The number of iterations of PGD in {\sc ObsAtk} is set to be five. 

% \hspace{-30em}
\begin{table}[h]
    \centering
    \caption{The average PSNR (in dB) results of DnCNN denoisers on the gray-scale and RGB datasets. 
    Four types of noise are used for evaluation, viz. Gaussian $\mcal N$ and Uniform $\mcal U$ random noise, and {\sc ObsAtk} with two different adversarial budgets. The energy-density of noise is bounded by $\hat \epsilon^2$.}
    \scalebox{0.85}{
    \begin{tabular}{cccccc}
    % \begin{tabular}{llllll}
    \toprule
    {Dataset} & $\hat \epsilon$ & $\mcal N$ & $\mcal U$ & Atk-\nicefrac{5}{255} & Atk-\nicefrac{7}{255} \\
    \hline
	\multirow{3}{*}{\small Set68}
	& \nicefrac{25}{255} &   29.16/0.02 & 29.15/0.01 & 24.26/0.12 & 23.12/0.10  \\
    & \nicefrac{15}{255} &   31.68/0.00 & 31.68/0/00 & 26.66/0.04 & 26.08/0.02  \\
% 	& \nicefrac{10}{255} &   33.84 & 33.85 & 29.22 & 28.61  \\
    \midrule
	\multirow{3}{*}{\small Set12}
	& \nicefrac{25}{255} &   30.39/0.01 & 30.41/0.01 & 24.32/0.18 & 22.96/0.13  \\
    & \nicefrac{15}{255} &   32.78/0.00 & 32.81/0.00 & 26.91/0.05 & 26.25/0.01  \\
% 	& \nicefrac{10}{255} &   34.72 & 34.71 & 29.51 & 28.58  \\
%    \bottomrule
	\midrule
	\multirow{3}{*}{\small BSD68}
	& \nicefrac{25}{255} &   31.25/0.11 & 31.17/0.11 & 27.44/0.08 & 26.08/0.06  \\
    & \nicefrac{15}{255} &   33.98/0.11 & 33.93/0.12 & 29.31/0.08 & 27.84/0.04  \\
% 	& \nicefrac{10}{255} &   36.32 & 36.30 & 30.85 & 29.46  \\
    \midrule
	\multirow{3}{*}{\small Kodak24}
	& \nicefrac{25}{255} &   32.20/0.13 & 32.13/0.14 & 27.87/0.08 & 26.37/0.07  \\
    & \nicefrac{15}{255} &   34.77/0.13 & 34.73/0.14 & 29.55/0.07 & 28.00/0.04  \\
% 	& \nicefrac{10}{255} &   36.90 & 36.89 & 30.98 & 29.55  \\
    \bottomrule
    \end{tabular}
    }
    \label{tab:attack-gray}
\end{table}{}

From Tables \ref{tab:attack-gray}, we observe that {\sc ObsAtk} clearly degrades the reconstruction performance of DIDs. In comparison to Gaussian or Uniform noisy images with the same noise levels, the recovered results from adversarial images are much worse in the sense of the PSNR. For example, when removing $\hat \epsilon=\nicefrac{15}{255}$ noisy images in Set12, the average PSNR of reconstructions from Gaussian noise can achieve 32.78 dB, whereas the PSNR drops to 26.25 dB when dealing with Atk-$\nicefrac{7}{255}$ adversarial images. We observe the consistent phenomenon that a normally-trained denoiser $f^{\epsilon}_{\theta}(\cdot)$ cannot effectively remove adversarial noise from visual results in Figure~\ref{fig:adv-noise}. 

\section{Robust and Generalizable Denoising via {\sc HAT}}
\label{sec:hat}
The previous section shows that existing deep denoisers are vulnerable to adversarial perturbations. 
To improve the adversarial robustness of deep denoisers, we propose an adversarial training method, hybrid adversarial training ({\sc HAT}), that uses original noisy images and their adversarial versions for training. 
Furthermore, we build a connection between the adversarial robustness of deep denoisers and their generalization capability to unseen types of noise. We show that {\sc HAT}-trained denoisers can effectively remove real-world noise without the need to leverage the real-world noisy data.
\subsection{Hybrid Adversarial Training}
\label{sec:hat-method}
AT has been proved to be a successful and universally applicable technique for robustifying deep neural networks. Most variants of AT are developed for the classification task specifically, such as TRADES \cite{zhang_theoretically_2019} and GAIRAT \cite{zhang_geometry-aware_2020}. Here, we propose an AT strategy, {\sc HAT}, for robust image denoising:
\begin{align}
	\min_{\theta\in \Theta} \E_{\rvx\sim P} & \E_{Q\sim \mcal U(\mcal Q^\epsilon)}  \E_{\rvv\sim Q}  \frac{1}{2} \Big(\frac{1}{1+\alpha}\|f_{\theta}^{\epsilon}(\rvy)-\rvx \|_2^2 \nonumber \\
	& + \frac{\alpha}{1+\alpha}\|f_{\theta}^{\epsilon}(\rvy)-f_{\theta}^{\epsilon}(\rvy') \|_2^2 \Big), 
	\label{eq:hybrid-at}
\end{align}
where $\rvy = \rvx+\rvv$ and $\rvy'=\rvy+\bm{\delta}^*$. Note that $\bm{\delta}^*$ is the adversarial perturbation obtained by solving {\sc ObsAtk} in Eq.~\eqref{eq:obsattack}.

As shown in Eq.~\eqref{eq:hybrid-at}, the loss function consists of two terms. The first term measures the distance between ground-truth images $\vx$ and reconstructions $f_{\theta}^{\epsilon}(\vy)$ from non-adversarial noisy images $\vy$, where $\vy$ contains noise $\vv$ sampled from a certain common distribution $Q$, such as Gaussian. This term encourages a good reconstruction performance of $f_{\theta}^{\epsilon}$ from common distributions. The second term is the distance between $f_{\theta}^{\epsilon}(\vy)$ and the reconstruction $f_{\theta}^{\epsilon}(\vy')$ from the adversarially perturbed version $\vy'$ of $\vy$. This term ensures that the reconstructions from any two noisy observations within a small neighborhood of $\vy$ have similar image qualities. Minimizing these two terms at the same time controls the worst-case reconstruction performance $\|f_{\theta}^{\epsilon}(\vy')-\vx\|$. 

The coefficient $\alpha$ balances the trade-off between reconstruction from common noise and the local continuity of $f_{\theta}^{\epsilon}$. When $\alpha$ equals zero, {\sc HAT} degenerates to normal training on common noise. The obtained denoisers fail to resist adversarial perturbations as shown in Section \ref{sec:obsattack}. When $\alpha$ is very large, the optimization gradually ignores the first term and completely aims for local smoothness. This may yield a trivial solution that $f_{\theta}^{\epsilon}$ always outputs a constant vector for any input. A proper value of $\alpha$ thus ensures a denoiser that performs well for common noise and the worst-case adversarial perturbations simultaneously. We perform an ablation study on the effect of $\alpha$ for the robustness enhancement and unseen noise removal in Appendix \ref{sec:apdx_ablation}.

To train a denoiser applicable to different levels of noise with an energy-density bounded by $\epsilon^2$, we randomly select a noise distribution $Q$ from a family of common distributions $\mcal Q^{\epsilon}$. $\mcal Q^{\epsilon}$ includes a variety of zero-mean distributions whose variance are bounded by $\epsilon^2$. For example, we define 
% $\mcal Q^{\epsilon}_{\mcal N} 
% = \{\mcal N(0,\sigma^2)| \sigma \sim \mcal U(0, \epsilon)\} 
$\mcal Q^{\epsilon}_{\mcal N} = \{ \mcal N(\bm{0}, \sigma^2 \mathbf{I}))|\sigma \sim\mcal U(0, \epsilon) \}$ for the experiments in the remaining of this paper.

\subsection{Robustness Enhancement via {\sc HAT}}
\label{sec:hat_for_robustness}

\def \SubFigWidth {0.3} % define a variable
\def \SubImgWidth {.95}
\begin{figure}[t!]
    \centering
    \begin{subfigure}{\SubFigWidth\linewidth}
        \centering
        \includegraphics[width=\SubImgWidth \linewidth]{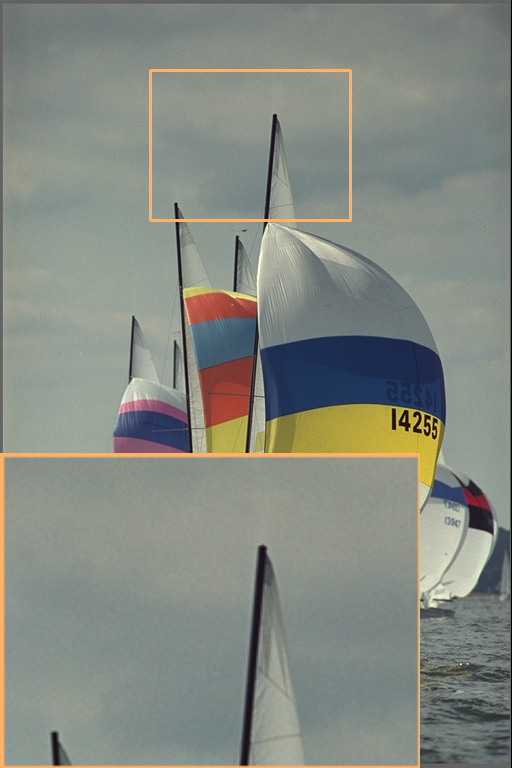}
        \caption{\scriptsize Ground-truth}
    \end{subfigure}
    \begin{subfigure}{\SubFigWidth\linewidth}
        \centering
        \includegraphics[width=\SubImgWidth \linewidth]{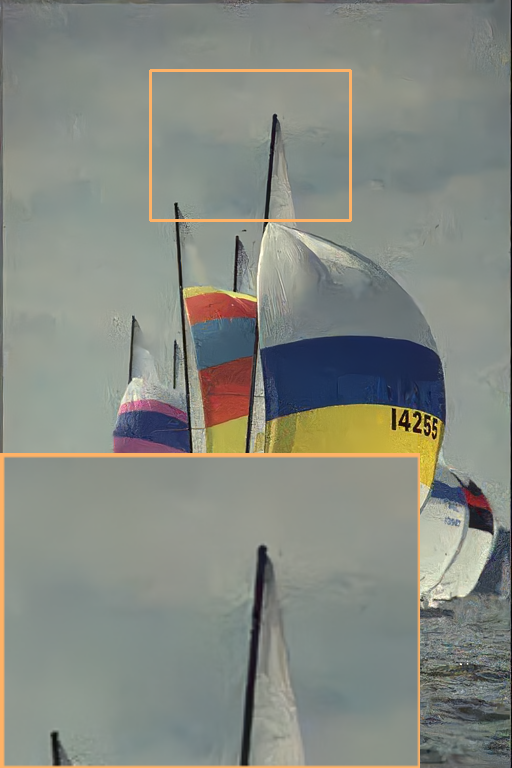}
        \caption{\scriptsize NT}
    \end{subfigure}
    \begin{subfigure}{\SubFigWidth\linewidth}
        \centering
        \includegraphics[width=\SubImgWidth \linewidth]{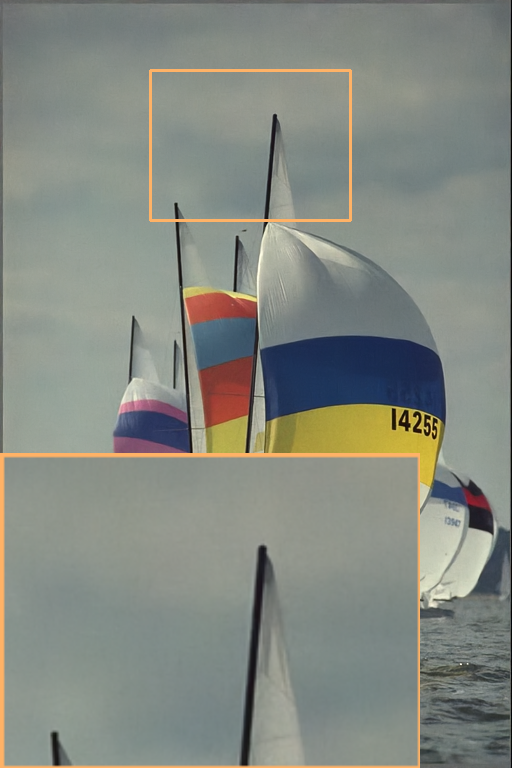}
        \caption{\scriptsize {\sc HAT}}
    \end{subfigure}
	\caption{
		From left to right are the ground-truth, the reconstruction of a normally-trained denoiser against attack, and the reconstruction of a {\sc HAT}-trained denoiser against attack.}
	\label{fig:hat-defense}
\end{figure}

We follow the same settings as those in Section \ref{sec:obsattack} for training and evaluating $\epsilon$-deep denoisers. The highest level of noise used for training is set to be $\epsilon=\nicefrac{25}{255}$. Noise is sampled from a set of Gaussian distributions $\mcal Q^{\epsilon}_{\mcal N}$.
We train deep denoisers with the {\sc HAT} strategy and set $\alpha$ to be $1$, and use one-step Atk-$\nicefrac{5}{255}$ to generate adversarially noisy images for training. We compare {\sc HAT} with normal training (NT) and the vanilla adversarial training (vAT) used in Choi \etal \shortcite{choi_deep_2021} that trains denoisers only with adversarial data. The results on Set68 and BSD68 are provided in this section. More results on Set12 and Kodak24 (in Tables \ref{tab:hat-set12} and \ref{tab:hat-kodak24}) are provided in Appendix \ref{sec:apdx_robustness_enhancement}. 

\begin{table}[h!]
    \centering
    \caption{The average PSNR (in dB) results of DnCNN-B denoisers on the gray-scale Set68 dataset. NT and {\sc HAT} are compared in terms of the noise removal of Gaussian noise and adversarial noise. We repeat the training for three times and report the mean and standard deviation (mean/std).}
    \scalebox{0.85}{
    \begin{tabular}{ccccccc}
    \toprule
    Method & $\hat \epsilon$ & $\mcal N$ & Atk-\nicefrac{3}{255} & Atk-\nicefrac{5}{255} & Atk-\nicefrac{7}{255} \\
    \hline
	\multirow{3}{*}{NT}
	& \nicefrac{25}{255} &   \thl{29.16}/0.02 & 26.20/0.07 & 24.26/0.12 & 23.12/0.10  \\
    & \nicefrac{15}{255} &   \thl{31.68}/0.00 & 27.98/0.05 & 26.66/0.04 & 26.08/0.02  \\
% 	& \nicefrac{10}{255} &   33.84 & 30.00 & 29.22 & 28.61  \\
	\midrule
	\multirow{3}{*}{vAT}
	& \nicefrac{25}{255} &   29.05/0.07 & 27.02/0.15 & 25.51/0.32 & 24.34/0.34 \\
    & \nicefrac{15}{255} &   31.53/0.09 & 28.74/0.16 & 27.43/0.19 & 26.68/0.15 \\
	\midrule
	\multirow{3}{*}{{\sc HAT}}
	& \nicefrac{25}{255} &   {28.88}/0.04 & \thl{27.48}/0.10 & \thl{26.40}/0.16 & \thl{25.32}/0.17 \\
    & \nicefrac{15}{255} &   {31.36}/0.03 & \thl{29.52}/0.01 & \thl{28.34}/0.03 & \thl{27.34}/0.03 \\
% 	& \nicefrac{10}{255} &   33.43 & 31.25 & 29.98 & 28.98 \\
    \bottomrule
    \end{tabular}
    }
    \label{tab:hat-set68}
\end{table}{}

From Tables \ref{tab:hat-set68} and \ref{tab:hat-bsd68}, we observe that {\sc HAT} obviously improves the reconstruction performance from adversarial noise in comparison to normal training. For example, on the Set68 dataset (Table \ref{tab:hat-set68}), when dealing with $\nicefrac{15}{255}$-level noise, the normally-trained denoiser achieves 31.68 dB for Gaussian noise removal, but the PSNR drops to 26.08 dB against Atk-\nicefrac{7}{255}. In contrast, the {\sc HAT}-trained denoiser achieves a PSNR of 27.34 dB (1.26 dB higher) against Atk-\nicefrac{7}{255} and maintains a PSNR of 31.36 dB for Gaussian noise removal. In Figure~\ref{fig:hat-defense}, we can see that when dealing with adversarially noisy images, the {\sc HAT}-trained denoiser can recover high-quality images while the normally-trained denoiser preserves noise patterns in the output. Besides, we observe that, similar to image classification tasks \cite{zhang_theoretically_2019},  AT-based methods ({\sc HAT} and vAT) robustify deep denoisers at the expense of the performance on non-adversarial data (Gaussian denoising). Nevertheless, the degraded reconstructions are still reasonably good  in terms of the PSNR. 
% Besides, we show that, in the following subsection, {\sc HAT} significantly enhances the generalization capability to unseen real-world noise.

\begin{table}[h!]
    \centering
    \caption{The average PSNR (in dB) results of DnCNN-C denoisers on the RGB BSD68 dataset.}
    \scalebox{0.85}{
    \begin{tabular}{ccccccc}
    \toprule
    Method & $\hat \epsilon$ & $\mcal N$ & Atk-\nicefrac{3}{255} & Atk-\nicefrac{5}{255} & Atk-\nicefrac{7}{255} \\
    \hline
	\multirow{3}{*}{NT}
	& \nicefrac{25}{255} &   \thl{31.25}/0.11 & 28.93/0.08 & 27.44/0.08 & 26.08/0.06  \\
    & \nicefrac{15}{255} &   \thl{33.98}/0.11 & 31.09/0.10 & 29.31/0.08 & 27.84/0.04  \\
% 	& \nicefrac{10}{255} &   36.32 & 32.87 & 30.85 & 29.46  \\
	\midrule
	\multirow{3}{*}{vAT}
	& \nicefrac{25}{255} &   30.64/0.02 & 28.81/0.03 & 27.67/0.01 & 26.64/0.03 \\
    & \nicefrac{15}{255} &   33.45/0.06 & 31.10/0.05 & 29.79/0.02 & 28.63/0.08 \\
    \midrule
	\multirow{3}{*}{{\sc HAT}}
	& \nicefrac{25}{255} &   {30.98}/0.03 & \thl{29.18}/0.03  & \thl{28.02}/0.02 & \thl{26.93}/0.04 \\
    & \nicefrac{15}{255} &   {33.67}/0.04 & \thl{31.38}/0.04  & \thl{30.03}/0.02 & \thl{28.80}/0.01 \\
% 	& \nicefrac{10}{255} &   35.83 & 33.04  & 31.51 & 30.21 \\
    \bottomrule
    \end{tabular}
    }
    \label{tab:hat-bsd68}
\end{table}{}

\begin{table*}[t!]
    \centering
    \caption{Comparison of different methods for denoising real-world noisy images in terms of PSNR (dB). We repeat the experiments of each denoising method for three times and report the  mean/standard deviation of PSNR values.
    }
    \scalebox{0.9}{
    \begin{tabular}{ccHcccccc}
    \toprule
    Dataset &  BM3D & BM3D & DIP & N2S(1) & NT & vAT & {\sc HAT} & N2C\\
    \hline
    % Fluorescence 	& 32.71 / 0.00 & - & 31.87 / 0.04 & 32.02 / 0.06 & 32.41 / 0.10 & 32.54 / 0.08 & \thl{33.32} / 0.03 & \tHL{34.88} / --  \\
    PolyU 	        & 37.40 / 0.00 & - & 36.08 / 0.01 & 35.37 / 0.15 & 35.86 / 0.01 & 36.77 / 0.00 & \thl{37.82} / 0.04 & -- / --    \\
    CC 		        & 35.19 / 0.00 & - & 34.64 / 0.06 & 34.33 / 0.14 & 33.56 / 0.01 & 34.49 / 0.10 & \thl{36.26} / 0.06 & -- / --  \\
    SIDD 	        & 25.65 / 0.00 & - & 26.89 / 0.02 & 26.51 / 0.03 & 27.20 / 0.70 & 27.08 / 0.28 & \thl{33.44} / 0.02 & \tHL{33.50} / 0.03  \\
    \bottomrule
    \end{tabular}
    }
    \label{tab:real-denoising}
\end{table*}{}

\subsection{Robustness Benefits Generalization to Unseen Noise}
\label{sec:hat_for_unseen}

It has been shown that denoisers that are normally trained on common synthetic noise fail to remove real-world noise induced by standard imaging procedures \cite{xu_multi-channel_2017,abdelhamed_high-quality_2018}. 
To train denoisers that can handle real-world noise, researchers have proposed several methods which can be roughly divided into two categories, namely \tit{dataset-based} denoising methods and \tit{single-image-based} denoising methods. High-performance dataset-based methods require a set of real noisy data for training, e.g., CBDNet requiring pairs of clean and noisy images \cite{guo_toward_2019} and Noise2Noise requiring multiple noisy observations of every single image \cite{lehtinen_noise2noise_2018}. 
However, a large number of paired data are not available in some applications, such as medical radiology and high-speed photography. 
To address this, single-image-based methods are proposed to remove noise by exploiting the correlation between signals across pixels and the independence between noise. This category of methods, such as DIP \cite{ulyanov_deep_2018} and N2S \cite{batson_noise2self_2019}, are adapted to various types of signal-independent noise, but they optimize the deep denoiser on each test image. The test-time optimization is extremely time-consuming, e.g., N2S needs to update a denoiser for \tit{thousands of iterations} to achieve good reconstruction performance. 

Here, we point out that {\sc HAT} is a promising framework to train a generalizable deep denoiser \tit{only with synthetic noise}. The resultant denoiser can be directly applied to perform  denoising for unseen noisy images in real-time. During training, {\sc HAT} first samples noise from common distributions (Gaussian) with noise levels from low to high. {\sc ObsAtk} then explores the $\rho$-neighborhood for each noisy image to search for a particular type of noise that degrades the denoiser the most. By ensuring the denoising performance of the worst-case noise, the resultant denoiser can deal with other unknown types of noise within the $\rho$-neighborhood as well. To train a robust denoiser that generalizes well to real-world noise, we need to choose a proper adversarial budget $\rho$. When $\rho$ is very small and close to zero, the {\sc HAT} reduces to normal training. When $\rho$ is very much larger than the norm of basic noise $\vv$, the adversarially noisy image may be visually unnatural because the adversarial perturbation $\bm{\delta}$ only satisfies the zero-mean constraint and is not guaranteed to be spatially uniformly distributed as other types of natural noise being. In practice, we set the value of $\rho$ of {\sc ObsAtk} to be $\nicefrac{5}{255}\cdot \sqrt{m}$, where $m$ denotes the size of image patches. The value of $\alpha$ of {\sc HAT} is kept unchanged as $2$.

\paragraph{Experimental Settings} We evaluate the generalization capability of {\sc HAT} on several real-world noisy datasets, including PolyU \cite{xu_real-world_2018}, CC \cite{xu_multi-channel_2017}, and SIDD \cite{abdelhamed_high-quality_2018}. 
PolyU, CC, and SIDD contain RGB images of common scenes in daily life. These images are captured by different brands of digital cameras and smartphones, and they contain various levels of noise by adjusting the ISO values. For the PolyU and CC, we use the clean images in BSD500 for training an adversarially robust $\epsilon$-denoiser with $\epsilon=\nicefrac{25}{255}$. We sample Gaussian noise from a set of distributions $\mcal Q^{\epsilon}_{\mcal N}$ and add the noise to clean images to craft noisy observations. {\sc HAT} trains the denoiser jointly with Gaussian noisy images and their adversarial versions.
For the SIDD, we use clean images in the SIDD-small set for training and test the denoisers on the SIDD-val set. The highest level of noise used for {\sc HAT} is set to be $\epsilon=\nicefrac{50}{255}$. In each case, we only use clean images for training denoisers without the need of real noisy images

\paragraph{Results} We compare {\sc HAT}-trained denoisers with the NT and vAT-trained ones. From Table \ref{tab:real-denoising}, we observe that {\sc HAT} performs much better than both competitors.
% The NT and vAT-trained ones cannot adapt to real noise removal. 
For example, on the SIDD-val dataset, the {\sc HAT}-trained denoiser achieves an average PSNR value of 33.44 dB that is 6.24 dB higher than the NT-trained one. We also compare {\sc HAT}-trained denoisers with single-image-based methods, including DIP, N2S, and the classical BM3D \cite{dabov_image_2007}. For DIP and N2S,\footnote{The officially released codes of DIP and N2S are used here. 
% \\ DIP: \url{https://github.com/DmitryUlyanov/deep-image-prior} \\N2S single-shot denoising: \url{https://github.com/czbiohub/noise2self}
} the numbers of iterations for each image are set to be 2,000 and 1,000, respectively. N2S works in two modes, namely single-image-based denoising and dataset-based denoising. Here, we use N2S in the single-image-based mode, denoted as N2S(1), due to the assumption that no real noisy data are available for training. We observe that {\sc HAT}-trained denoisers consistently outperform these baselines. Visual comparisons are provided in Appendix \ref{sec:apdx_vis_real}. Besides, since the SIDD-small provides a set of real noisy and ground-truth pairs, we train a denoiser, denoted as Noise2Clean (N2C), with these paired data and use the N2C denoiser as the oracle for comparison. We observe that {\sc HAT}-trained denoisers are comparable to the N2C one for denoising images in SIDD-val (a PSNR of 33.44dB vs 33.50dB).

\section{Conclusion}
Normally-trained deep denoisers are vulnerable to adversarial attacks. {\sc HAT} can effectively robustify deep denoisers and boost their generalization capability to unseen real-world noise. In the future, we will extend the adversarial-training framework to other image restoration tasks, such as deblurring. We aim to develop a generic AT-based robust optimization framework to train deep models that can recover clean images from unseen types of degradation.

%\clearpage
\section{Acknowledgments}
HY and VYFT are funded by a Singapore National Research Foundation (NRF) Fellowship (R-263-000-D02-281). \\
JZ was supported by JST ACT-X Grant Number JPMJAX21AF. \\
MS was supported by JST CREST Grant Number JPMJCR18A2.

\bibliographystyle{named}
\bibliography{ijcai22}

%\clearpage
\onecolumn
\appendix
\begin{center}
    \huge \textbf{Appendices}
\end{center}

\counterwithin{figure}{section}
\counterwithin{table}{section}

\section{Two-step Projection}
\label{sec:thm1-proof}
\setcounter{theorem}{0}
\begin{theorem} 
	For any arbitrary vector $\vdelta \in \sR^m$, its projection onto the region defined by the intersection of the  norm-bounded and zero-mean constraints  is equivalent to the projection first onto the zero-mean hyperplane followed by the projection onto the $\rho$-ball ($\rho>0$), i.e., 
	\begin{equation}
		\proj_{A\cap B}(\vdelta) = \proj_B(\proj_A(\vdelta)),
		\label{eq:thm}
	\end{equation}
	 where
\begin{subequations}
\begin{align}
	A & = \big\{\vz\in \sR^m  \,\big|\, \vn^{\top}\vz = 0 \big\}, \\
	B & =\big\{\vz\in \sR^m  \,\big|\, \|\vz\|_2^2 \leq \rho^2\big\},
\end{align}
\end{subequations}
and  $\vn = [1,1, \ldots, 1]^\top$.
\end{theorem}

\begin{figure}[h!]
	\centering
	\includegraphics[width=.55\linewidth]{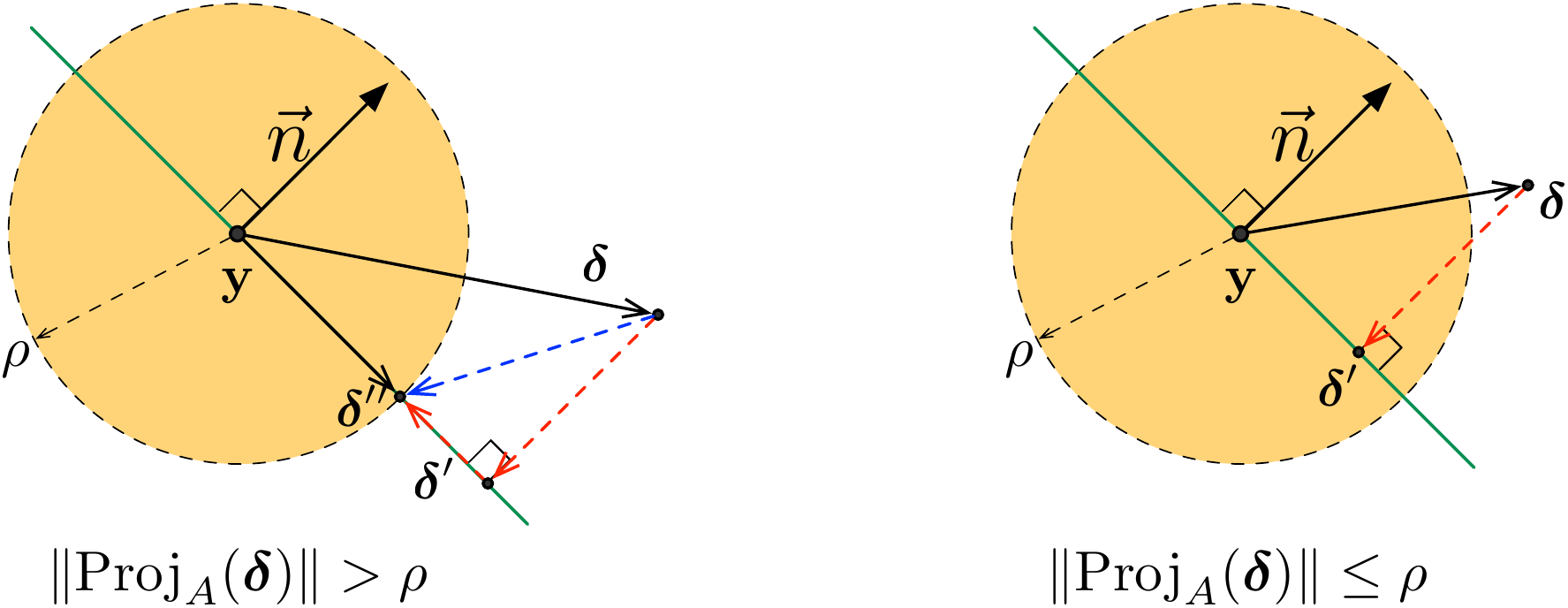}
	\caption{Illustration of Theorem 1. In the case of $\|\proj_A(\vdelta)\| > \rho$, the red-dot lines show that the perturbation $\vdelta$ is projected onto the region defined by the zero-mean and $\rho$-ball constraints sequentially. The blue-dot line shows the exact projection of $\vdelta$ on to $A\cap B$.}
\end{figure}

\paragraph{Proof} 
Let us consider the RHS of Eq. \eqref{eq:thm} first. It is easy to derive the projections onto $A$ and $B$ seperately: 
\begin{subequations}
\begin{align}
	\proj_A(\vdelta) & = \vdelta - \frac{\vn^\top \vdelta }{\|{\vn}\|^2_2} {\vn}, \label{eq:apdx-proj-zero-mean} \\
	\proj_B(\vdelta) & = \min\Big(\frac{\rho}{\|\vdelta\|_2}, 1\Big) \text{~} \vdelta. \label{eq:apdx-proj-norm}
\end{align}
\label{eq:apdx-two-step-projection}
\end{subequations}
Thus, we have
\begin{equation}
  \proj_B(\proj_A(\vdelta)) =
    \begin{cases}
      \proj_A(\vdelta), 
      	& \text{if $\|\proj_A(\vdelta)\| \leq \rho$};\\
      \frac{\rho}{\|\proj_A(\vdelta)\|_2} \proj_A(\vdelta), 
      	& \text{if $\|\proj_A(\vdelta)\| > \rho$.}
    \end{cases}
    \label{eq:two-step-projection-sol}   
\end{equation}

Now let us consider the LHS of Eq.~\eqref{eq:thm}. The projection onto $A\cap B$ can be formulated as the solution of the following convex optimization problem:
\begin{equation}
	\min_{\vz} \frac{1}{2}\|\vdelta -\vz\|^2_2, \quad
	\text{s.t.} \quad \vn^{\top} \vz = 0, \quad
	 \|\vz\|^2_2 \leq \rho^2,
\label{eq:proj-jointly-opt}
\end{equation}
where $\vz \in \sR^m$. We can write the Lagrangian, $L:\sR^m \times \sR \times \sR \to \sR$, associated with the problem \eqref{eq:proj-jointly-opt} as 
\begin{equation}
	L(\vz, \lambda, \nu) = \frac{1}{2}\|\vdelta -\vz\|^2_2 + \lambda (\|\vz^*\|^2_2 - \rho^2) + \nu \vn^{\top} \vz.
\end{equation}
%with a domain of $A\cap B) \times \sR \times \sR$.
Since there exists an $\vz \in \sR^m$, e.g., $\vz=[0,\ldots,0]^{\top} \in \sR^m$, such that $\vn^{\top}\vz=0$ and $\|\vz\|^2_2 < \rho^2$, the problem \eqref{eq:proj-jointly-opt} is strictly feasible, i.e., it satisfies the Slater's condition \cite{boyd2004convex}. Besides, the objective and the constraints are all differentiable, thus the KKT conditions in Eq.~\eqref{eq:proj-jointly-opt-kkt} provide necessary and sufficient conditions for optimality.
\begin{subequations}
\begin{align}
	\|\vz^*\|^2_2 - \rho^2 \leq 0, \\
	\vn^{\top}\rz^* = 0, \\
	\lambda \geq 0,\\
	\lambda (\|\vz^*\|^2_2 - \rho^2) = 0, \\
	\frac{\partial L}{\partial \vz} = (1+2\lambda) \vz^* - \vdelta + \nu \vn= 0.
\end{align}
\label{eq:proj-jointly-opt-kkt}
\end{subequations}

We obtain the optimal solution by considering the following two cases separately, i.e., $\|\proj_A(\vdelta)\| \leq \rho$ and $\|\proj_A(\vdelta)\| > \rho$. 

~\\
\textbf{Case-(1)}: $\|\proj_A(\vdelta)\| > \rho$. \\
If $\lambda>0$, then Eq. \eqref{eq:proj-jointly-opt-kkt} reduces to the following equation:
\begin{subequations}
\begin{align}
	\vn^{\top}\rz^* = 0, \\
	\|\vz^*\|^2_2 - \rho^2 = 0, \\
	(1+2\lambda) \vz^* - \vdelta + \nu \vn= 0.
\end{align}
\end{subequations}
We can easily solve these equations and obtain that
\begin{subequations}
\begin{align}
	(1+2\lambda) &= \proj_A(\vdelta) / \rho, \\
	\nu &= \vn^{\top}\vdelta / m, \\
	\vz^* &= \frac{\rho}{\|\proj_A(\vdelta)\|_2} \proj_A(\vdelta). 
\end{align}
\end{subequations}
If $\lambda=0$, then Eq. \eqref{eq:proj-jointly-opt-kkt} reduces to the following  set of equations:
\begin{subequations}
\begin{align}
	\|\vz^*\|^2_2 - \rho^2 &\leq 0, \\
	\vn^{\top}\rz^* &= 0 \label{eqn:bottom2a}\\
	\vz^* - \vdelta + \nu \vn&= 0.  \label{eqn:bottom2b}
\end{align}
\end{subequations}
According to \eqref{eqn:bottom2a} and \eqref{eqn:bottom2b}, we obtain that $\vz^* = \vdelta - \frac{\vn^{\top}\vdelta}{m}\vn = \proj_A(\vdelta) $ with a norm strictly larger than $\rho$, which contradicts the constraint $\|\vz^*\|^2_2 \leq \rho^2$. Thus, for the case of  $\|\proj_A(\vdelta)\| > \rho$, we have that $\vz^* = \frac{\rho}{\|\proj_A(\vdelta)\|_2} \proj_A(\vdelta)$ which is equal to $\proj_B(\proj_A(\vdelta))$ in Eq. \eqref{eq:two-step-projection-sol}.

~\\
\textbf{Case-(2)}: $\|\proj_A(\vdelta)\| \leq \rho$. \\ 
Since $\|\proj_A(\vdelta)\| \leq \rho$ and $\|\proj_A(\vdelta)\| \in A$, we have $\proj_A(\vdelta) \in A\cap B$. For any other point $\vz' \in A\cap B$ and $\vz' \neq \proj_A(\vdelta)$, we have $\|\vdelta-\vz'\| > \|\vdelta-\proj_A(\vdelta)\| $, where the strict inequality holds because $A$ is the set of points from a hyperplane. Thus, $\vz' $ is not the $ \proj_{A\cap B}(\vdelta)$. Therefore, $\proj_{A\cap B}(\vdelta) = \proj_A(\vdelta) = \proj_B(\proj_A(\vdelta))$.

~\\
In summary, we show that $\proj_{A\cap B}(\vdelta) = \proj_B(\proj_A(\vdelta))$ for any arbitrary $\vdelta \in \sR^m$.

\section{Experiments of Robustness Enhancement on Set12 and Kodak24}
We compare the robustness of deep denoisers trained via three strategies, i.e., NT, vAT and HAT. The results on Set 12 and Kodak24 are provided in Table \ref{tab:hat-set12} and Table \ref{tab:hat-kodak24} respectively.
We observe that HAT can effectively robustify deep denoisers. The reconstruction quality of HAT-trained denoisers from adversarially noisy images is clearly better than that of the NT and vAT-trained ones.

\label{sec:apdx_robustness_enhancement}
\begin{table}[h!]
    \centering
    \caption{The average PSNR (in dB) results of DnCNN-B denoisers on the gray-scale Set12 dataset. }
    \scalebox{.9}{
    \begin{tabular}{ccccccc}
    \toprule
    Training & $\hat \epsilon$ & $\mcal N$ & Atk-\nicefrac{3}{255} & Atk-\nicefrac{5}{255} & Atk-\nicefrac{7}{255} \\
    \hline
	\multirow{3}{*}{NT}
	& \nicefrac{25}{255} &   \thl{30.39}/0.01 & 26.51/0.14 & 24.32/0.18 & 22.96/0.13  \\
    & \nicefrac{15}{255} &   \thl{32.78}/0.00 & 28.50/0.08 & 26.91/0.05 & 26.25/0.01  \\
% 	& \nicefrac{10}{255} &   34.72 & 30.50 & 29.51 & 28.58  \\
	\midrule
	\multirow{3}{*}{vAT}
	& \nicefrac{25}{255} &   30.25/0.08 & 27.56/0.06 & 25.82/0.04 & 24.33/0.04 \\
    & \nicefrac{15}{255} &   32.63/0.09 & 29.37/0.17 & 27.83/0.15 & 26.91/0.08 \\
    \midrule
	\multirow{3}{*}{HAT}
	& \nicefrac{25}{255} &   30.01/0.06 & \thl{27.96}/0.15 & \thl{26.46}/0.20 & \thl{25.13}/0.19 \\
    & \nicefrac{15}{255} &   32.47/0.04 & \thl{29.95}/0.03 & \thl{28.45}/0.04 & \thl{27.20}/0.03 \\
% 	& \nicefrac{10}{255} &   34.41 & 31.61 & 30.14 & 28.95 \\
    \bottomrule
    \end{tabular}
    }
    \label{tab:hat-set12}
\end{table}{}

\begin{table}[h!]
    \centering
    \caption{The average PSNR (in dB) results of DnCNN-C denoisers on the RGB Kodak24 dataset.}
    \scalebox{.9}{
    \begin{tabular}{ccccccc}
    \toprule
    Training & $\hat \epsilon$ & $\mcal N$ & Atk-\nicefrac{3}{255} & Atk-\nicefrac{5}{255} & Atk-\nicefrac{7}{255} \\
    \hline
	\multirow{3}{*}{NT}
	& \nicefrac{25}{255} &   \thl{32.20}/0.13 & 29.57/0.09 & 27.87/0.08 & 26.37/0.07  \\
    & \nicefrac{15}{255} &   \thl{34.77}/0.13 & 31.54/0.11 & 29.55/0.07 & 28.00/0.04  \\
% 	& \nicefrac{10}{255} &   36.90 & 33.11 & 30.98 & 29.55  \\
	\midrule
	\multirow{3}{*}{vAT}
	& \nicefrac{25}{255} &   31.44/0.01 & 29.41/0.05 & 28.13/0.06 & 26.98/0.02 \\
    & \nicefrac{15}{255} &   34.14/0.08 & 31.53/0.11 & 30.06/0.08 & 28.78/0.06 \\
    \midrule
	\multirow{3}{*}{HAT}
	& \nicefrac{25}{255} &   31.83/0.04 & \thl{29.85}/0.02  & \thl{28.56}/0.02 & \thl{27.34}/0.05 \\
    & \nicefrac{15}{255} &   34.36/0.06 & \thl{31.84}/0.05  & \thl{30.37}/0.02 & \thl{29.05}/0.01 \\
% 	& \nicefrac{10}{255} &   36.30 & 33.31  & 31.69 & 30.34 \\
    \bottomrule
    \end{tabular}
    }
    \label{tab:hat-kodak24}
\end{table}{}

%\vspace{10em}
\vfill
\section{Ablation study}
\label{sec:apdx_ablation}

\subsection{Effect of $\alpha$ on Robustness Enhancement and Generalization to Real-world noise}

Here, we evaluate the effect of $\alpha$ in HAT on the adversarial robustness and the generalization capability to real-world noise. We train deep denoisers on the RGB BSD500 (except 68 images for test) dataset. The obtained denoisers are tested on the BSD68 dataset for Gaussian and adversarial noise removal. The generalization capability is evaluated on two datasets of real-world noisy images, i.e., PolyU and CC. Experimental settings follow those in Section \ref{sec:hat_for_robustness}.

\begin{figure}[h!]
	\centering
	\includegraphics[width=.6\linewidth]{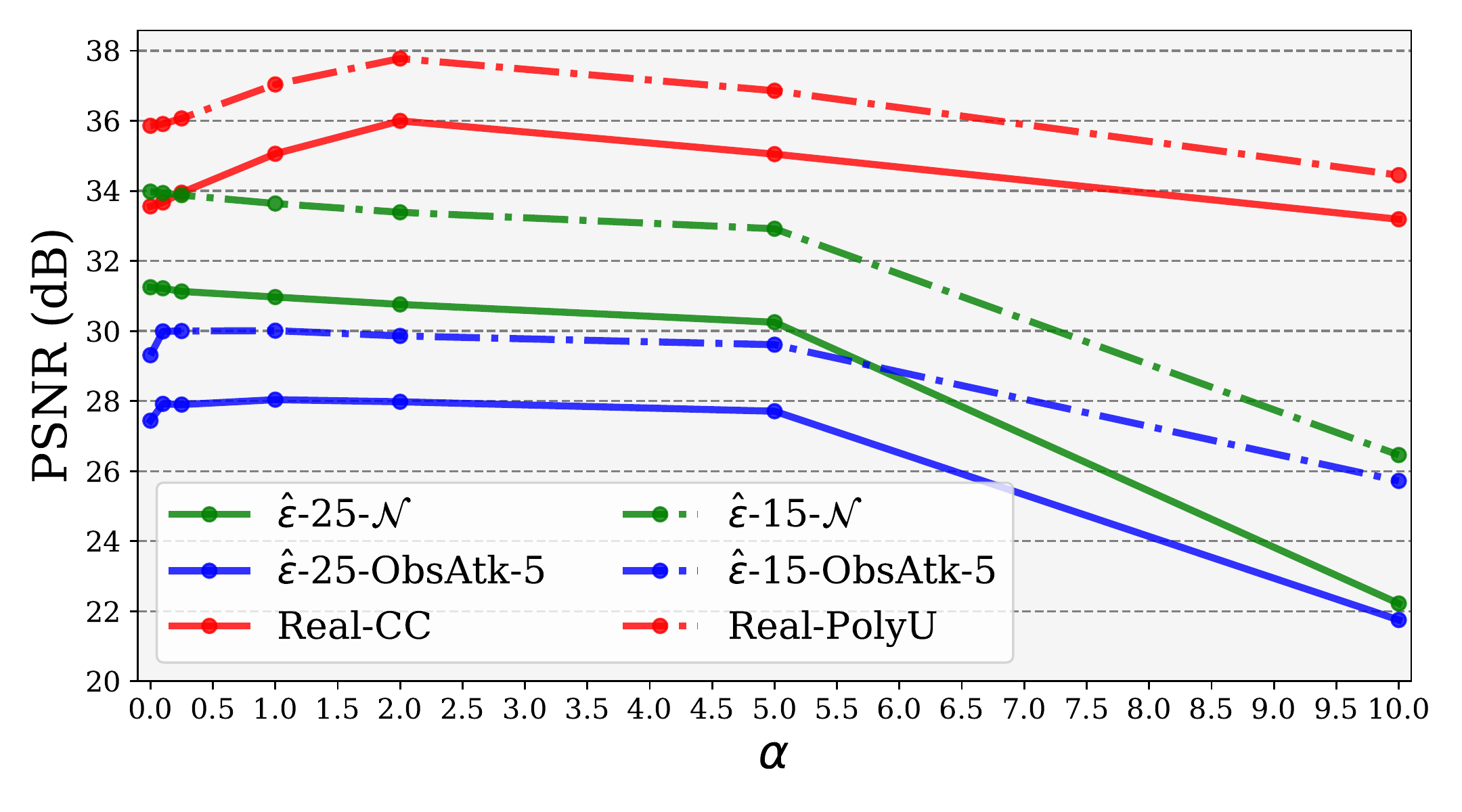}
	\caption{Ablation study on the effect of $\alpha$ in HAT. \textcolor{Green}{Green lines} show the denoising results on non-adversarial noise sampled from common distributions. The legend $\hat \epsilon$-w-$\mcal N$ denotes the Gaussian noise ($\sigma=\nicefrac{w}{255}$) with a energy-density bounded by $\hat \epsilon^2=\nicefrac{w^2}{255^2}$. \textcolor{Blue}{Blue lines} show that denoising results on adversarially perturbed noisy images. $\hat \epsilon$-w-ObsAtk-5 denotes the adversarial noise crafted by ObsAtk-5 with a energy-density bounded by $\hat \epsilon^2$. \textcolor{Red}{Red lines} show the denoising results on real-world noisy images.}
	\label{fig:apdx-ablation-alpha}
\end{figure}

Figure \ref{fig:apdx-ablation-alpha} corroborates the analysis in Section \ref{sec:hat-method} that the coefficient $\alpha$ balances the trade-off between reconstruction from common noise and the adversarial robustness. We also find that the generalization capability to real-world noise is correlated to the adversarial robustness. Specifically, good adversarial robustness usually implies good generalization to real-world noise. In Figure \ref{fig:apdx-ablation-alpha}, the best robustness and the best performance on real-world noise appear around $\alpha=1$ or $2$. When $\alpha$ is too large or too small, the robustness and generalization worsen. For the noise sampled from Gaussian distributions, increasing $\alpha$ degrades the denoising performance. In summary, we set $\alpha$ to $1$ or $2$ to achieve a good balance between the denoising performance on common noise and the adversarial robustness as well as real-world generalization.

\subsection{Effect of $\rho$ on Generalization to Real-world Noise}

Here, we evaluate the effect of $\rho$ used in HAT on the generalization capability to real-world noise. We train deep denoisers on the RGB BSD500 (except 68 images for test) dataset and evaluate the generalization capability on two real-world datasets, namely PolyU and CC. The $\alpha$ is set to be $2$. The adversarial budget $\rho$ of ObsAtk-$\nicefrac{\rho}{\sqrt{m}}$, that generates adversarially noisy images for HAT, is set to be values from $[0, \sqrt{m}, 3\sqrt{m},\ldots, 11\sqrt{m}]$ for comparison, where $m$ denotes the size of images. Other experimental settings follow those in Section \ref{sec:hat_for_robustness}.

\begin{figure}[h!]
	\centering
 	\includegraphics[width=.6\linewidth]{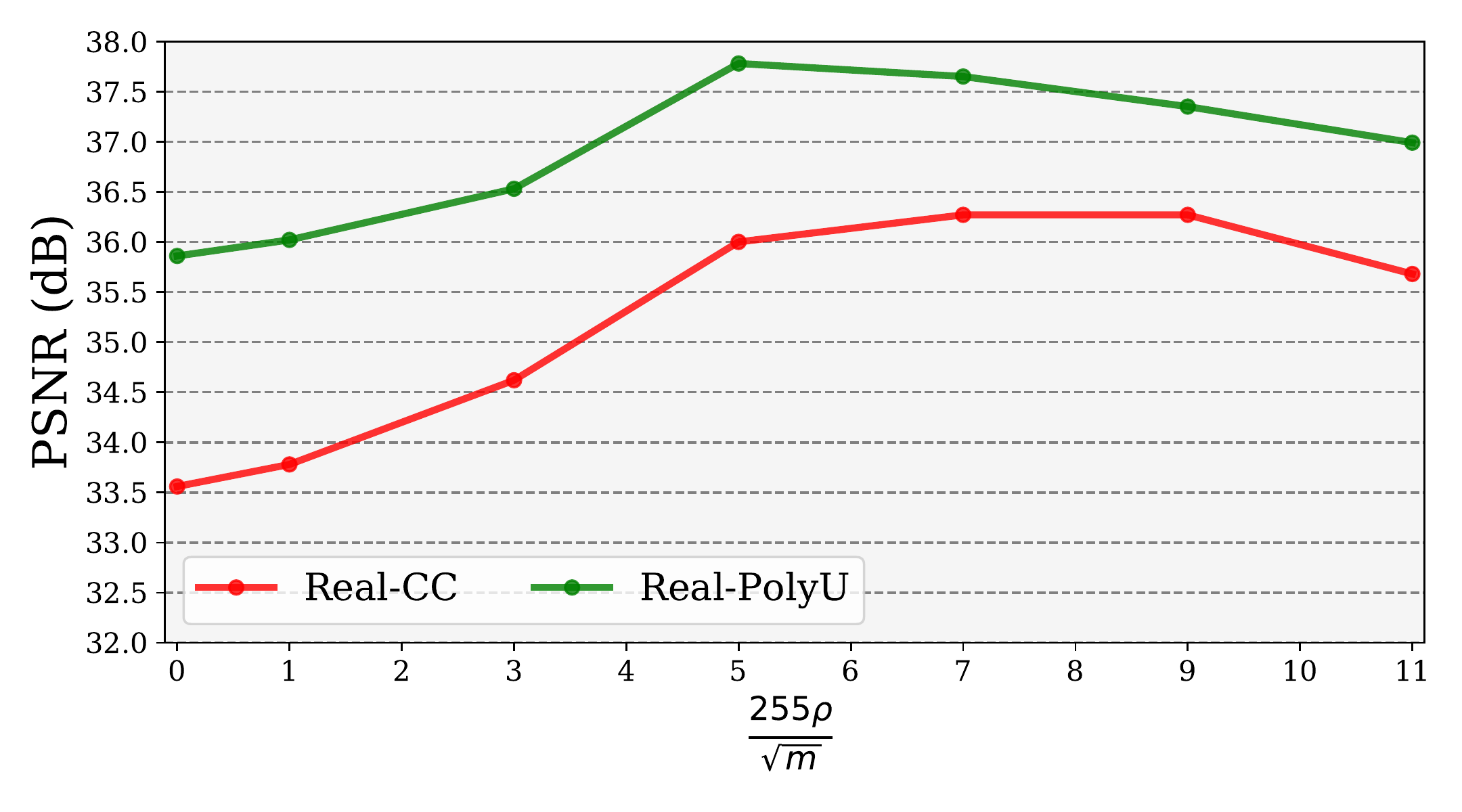}
	\caption{Ablation study on the effect of $\rho$ in HAT.}
	\label{fig:apdx-ablation-rho}
\end{figure}

Figure \ref{fig:apdx-ablation-rho} corroborates the analysis in Section \ref{sec:hat_for_unseen}.
%To train a robust denoiser that generalizes well to real-world noise, we need to choose a proper adversarial budget $\rho$. 
When $\rho$ is very small and close to zero, the HAT reduces to normal training. The resultant denoisers cannot effectively remove real-world noise. When $\rho$ is very much larger than the norm of basic noise $\vv$, the statistics of adversarial noise may be very unnatural because the adversarial perturbation $\bm{\delta}$ might concentrate on a certain region, like edges or texture, and not be spatially uniformly distributed as other types of natural noise being. We can see that, when $\rho > \nicefrac{7}{255} \sqrt{m}$, the denoising performance on real-world datasets starts to decrease. In practice, we set the value of $\rho$ of ObsAtk to be $\nicefrac{5}{255}\cdot \sqrt{m}$ to train generalizable denoisers.

\section{Visual results of real-world noise removal}
\label{sec:apdx_vis_real}

We show the denoising results on SIDD-val set in Figure \ref{fig:hat-sidd}. We observe that HAT-trained denoiser can effectively remove the real-world noise while the normally-trained one retains much noise in the reconstructions. Besides, the HAT-trained denoiser outperforms other baseline methods and produces much cleaner results. 
%Although the reconstructions of DIP and N2S look visually cleaner than the results of NT and vAT-trained denoisers, DIP and N2S cause the mean-shift in the output and thus produce worse reconstructions in terms of the numerical PSNR (refer to Table~\ref{tab:real-denoising} in the main article).

\def \SubFigWidth {0.12} % define a variable
\def \SubImgWidth {.95}
\begin{figure*}[h!]
    \centering
    \begin{subfigure}{\SubFigWidth\linewidth}
        \centering
        \includegraphics[width=\SubImgWidth \linewidth]{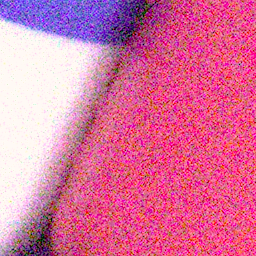}
    \end{subfigure}
    \begin{subfigure}{\SubFigWidth\linewidth}
        \centering
        \includegraphics[width=\SubImgWidth \linewidth]{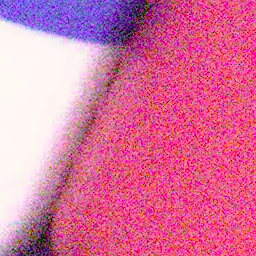}
    \end{subfigure}
    \begin{subfigure}{\SubFigWidth\linewidth}
        \centering
        \includegraphics[width=\SubImgWidth \linewidth]{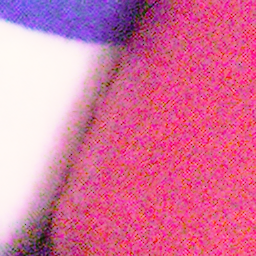}
    \end{subfigure}
    \begin{subfigure}{\SubFigWidth\linewidth}
        \centering
        \includegraphics[width=\SubImgWidth \linewidth]{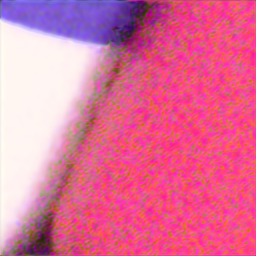}
    \end{subfigure}
    \begin{subfigure}{\SubFigWidth\linewidth}
        \centering
        \includegraphics[width=\SubImgWidth \linewidth]{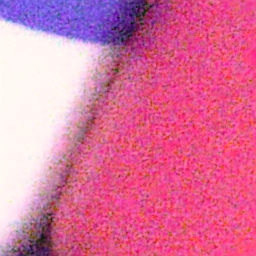}
    \end{subfigure}
    \begin{subfigure}{\SubFigWidth\linewidth}
        \centering
        \includegraphics[width=\SubImgWidth \linewidth]{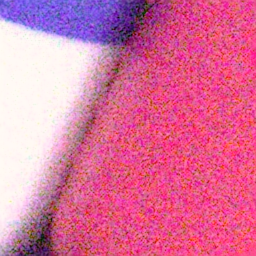}
    \end{subfigure}
    \begin{subfigure}{\SubFigWidth\linewidth}
        \centering
        \includegraphics[width=\SubImgWidth \linewidth]{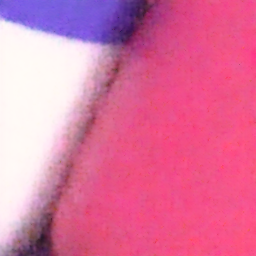}
    \end{subfigure}
    \begin{subfigure}{\SubFigWidth\linewidth}
        \centering
        \includegraphics[width=\SubImgWidth \linewidth]{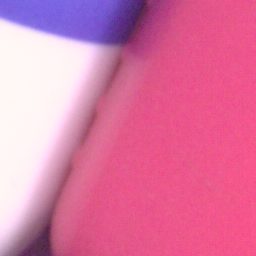}
    \end{subfigure}

    \begin{subfigure}{\SubFigWidth\linewidth}
        \centering
        \includegraphics[width=\SubImgWidth \linewidth]{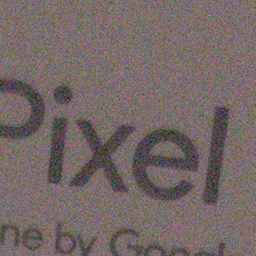}
    \end{subfigure}
    \begin{subfigure}{\SubFigWidth\linewidth}
        \centering
        \includegraphics[width=\SubImgWidth \linewidth]{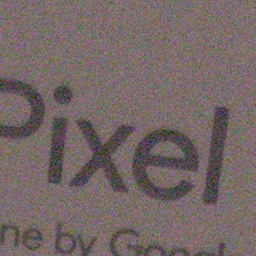}
    \end{subfigure}
    \begin{subfigure}{\SubFigWidth\linewidth}
        \centering
        \includegraphics[width=\SubImgWidth \linewidth]{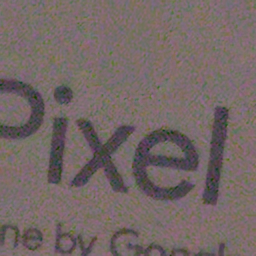}
    \end{subfigure}
    \begin{subfigure}{\SubFigWidth\linewidth}
        \centering
        \includegraphics[width=\SubImgWidth \linewidth]{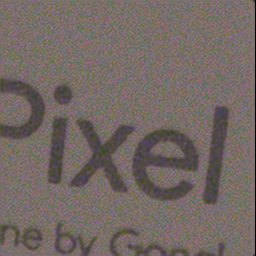}
    \end{subfigure}
    \begin{subfigure}{\SubFigWidth\linewidth}
        \centering
        \includegraphics[width=\SubImgWidth \linewidth]{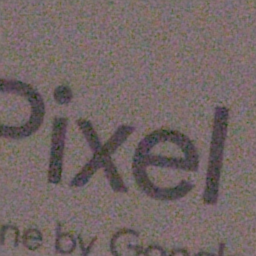}
    \end{subfigure}
    \begin{subfigure}{\SubFigWidth\linewidth}
        \centering
        \includegraphics[width=\SubImgWidth \linewidth]{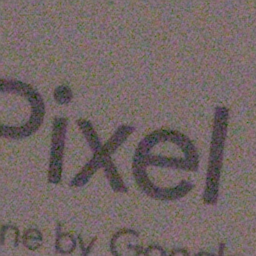}
    \end{subfigure}
    \begin{subfigure}{\SubFigWidth\linewidth}
        \centering
        \includegraphics[width=\SubImgWidth \linewidth]{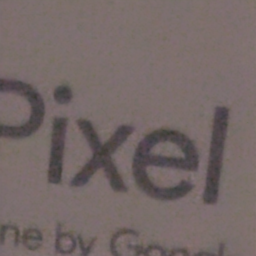}
    \end{subfigure}
    \begin{subfigure}{\SubFigWidth\linewidth}
        \centering
        \includegraphics[width=\SubImgWidth \linewidth]{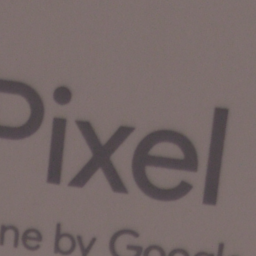}
    \end{subfigure}

    \begin{subfigure}{\SubFigWidth\linewidth}
        \centering
        \includegraphics[width=\SubImgWidth \linewidth]{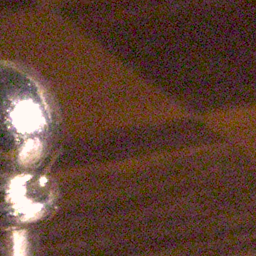}
    \end{subfigure}
    \begin{subfigure}{\SubFigWidth\linewidth}
        \centering
        \includegraphics[width=\SubImgWidth \linewidth]{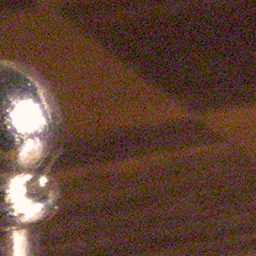}
    \end{subfigure}
    \begin{subfigure}{\SubFigWidth\linewidth}
        \centering
        \includegraphics[width=\SubImgWidth \linewidth]{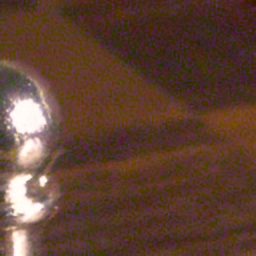}
    \end{subfigure}
    \begin{subfigure}{\SubFigWidth\linewidth}
        \centering
        \includegraphics[width=\SubImgWidth \linewidth]{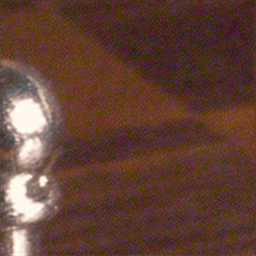}
    \end{subfigure}
    \begin{subfigure}{\SubFigWidth\linewidth}
        \centering
        \includegraphics[width=\SubImgWidth \linewidth]{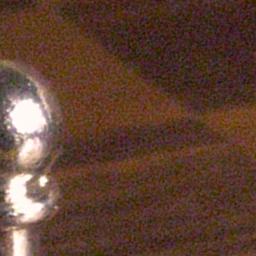}
    \end{subfigure}
    \begin{subfigure}{\SubFigWidth\linewidth}
        \centering
        \includegraphics[width=\SubImgWidth \linewidth]{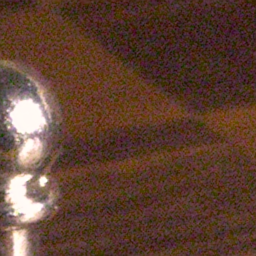}
    \end{subfigure}
    \begin{subfigure}{\SubFigWidth\linewidth}
        \centering
        \includegraphics[width=\SubImgWidth \linewidth]{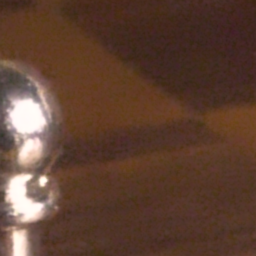}
    \end{subfigure}
    \begin{subfigure}{\SubFigWidth\linewidth}
        \centering
        \includegraphics[width=\SubImgWidth \linewidth]{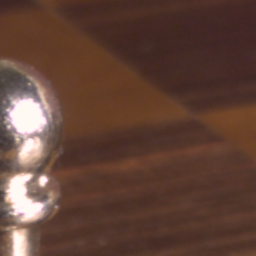}
    \end{subfigure}

    \begin{subfigure}{\SubFigWidth\linewidth}
        \centering
        \includegraphics[width=\SubImgWidth \linewidth]{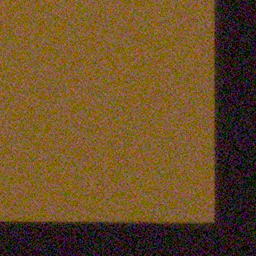}
        \caption{\scriptsize Noisy}
    \end{subfigure}
    \begin{subfigure}{\SubFigWidth\linewidth}
        \centering
        \includegraphics[width=\SubImgWidth \linewidth]{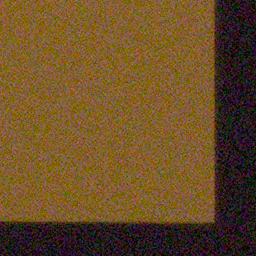}
        \caption{\scriptsize BM3D}
    \end{subfigure}
    \begin{subfigure}{\SubFigWidth\linewidth}
        \centering
        \includegraphics[width=\SubImgWidth \linewidth]{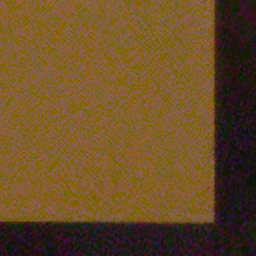}
        \caption{\scriptsize DIP}
    \end{subfigure}
    \begin{subfigure}{\SubFigWidth\linewidth}
        \centering
        \includegraphics[width=\SubImgWidth \linewidth]{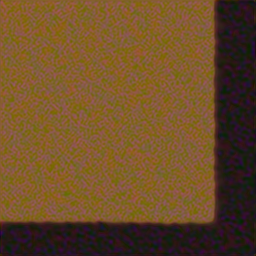}
        \caption{\scriptsize N2S}
    \end{subfigure}
    \begin{subfigure}{\SubFigWidth\linewidth}
        \centering
        \includegraphics[width=\SubImgWidth \linewidth]{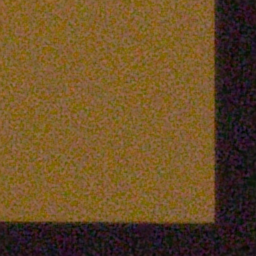}
        \caption{\scriptsize NT}
    \end{subfigure}
    \begin{subfigure}{\SubFigWidth\linewidth}
        \centering
        \includegraphics[width=\SubImgWidth \linewidth]{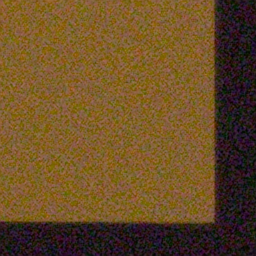}
        \caption{\scriptsize vAT}
    \end{subfigure}
    \begin{subfigure}{\SubFigWidth\linewidth}
        \centering
        \includegraphics[width=\SubImgWidth \linewidth]{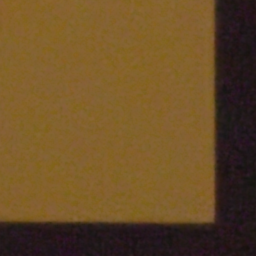}
        \caption{\scriptsize HAT}
    \end{subfigure}
    \begin{subfigure}{\SubFigWidth\linewidth}
        \centering
        \includegraphics[width=\SubImgWidth \linewidth]{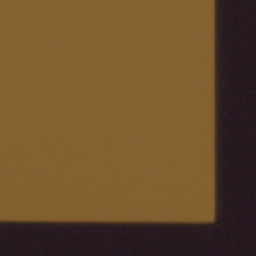}
        \caption{\scriptsize Ground-truth}
    \end{subfigure}
	\caption{
	Comparison of different denoisers for denoising SIDD-val set. From left to right are the input noisy image, reconstructions of different denoisers including BM3D, DIP, N2S, NT-trained DnCNN, vAT-trained DnCNN, and HAT-trained DnCNN. We can see that the HAT-trained denoiser performs the best in comparison to other baseline methods.
		}
	\label{fig:hat-sidd}
\end{figure*}

\end{document}